\documentclass[aps,prd,twocolumn,preprintnumbers,nofootinbib]{revtex4}
\usepackage{amsfonts,amssymb}
\usepackage{graphicx}

\let\bi=\bibitem
\newcommand{\be}[1]{\begin{equation}\label{#1}}
\newcommand{\ee}{\end{equation}}
\newcommand{\bea}[1]{\begin{eqnarray}\label{#1}}
\newcommand{\eea}{\end{eqnarray}}

\newcommand{\bZ}{\mathbb{Z}}
\newcommand{\bZZ}{\bZ_2\times\bZ_2}

\newcommand{\rep}[1]{\mathbf{#1}}

\newcommand{\fivebar}{\rep{\bar{5}}}
\newcommand{\ten}{\rep{10}}
\newcommand{\fifteen}{\rep{15}}
\newcommand{\Sym}{\mathrm{Sym}}
\newcommand{\Anti}{\mathrm{Anti}}

\begin{document}

\title{Statistics of SU(5) D-Brane Models on a Type II Orientifold}
\author{Florian Gmeiner$^{1,2,}$\footnote{flo AT mppmu.mpg.de}, Maren Stein$^{2,}$\footnote{mstein AT theorie.physik.uni-muenchen.de}}
\affiliation{%
${}^1\!\!\!$ Max-Planck-Institut f\"ur Physik, F\"ohringer Ring 6, 80805 M\"unchen, Germany\\
${}^2\!\!\!$ Arnold-Sommerfeld-Center for Theoretical Physics, Department f\"ur Physik, Ludwig-Maximilians-Universit\"at  M\"unchen, Theresienstra{\ss}e 37, 80333 M\"unchen, Germany}
\preprint{LMU-ASC 10/06\ \ \ \ MPP-2006-18\ \ \ \ hep-th/0603019}
\date{March 02, 2006}

\begin{abstract}
We perform a statistical analysis of models with SU(5) and flipped SU(5) gauge
group in a type II orientifold setup.
We investigate the distribution and correlation of properties of these models,
including the number of generations and the hidden sector gauge group.
Compared to the recent analysis hep-th/0510170 of models with a standard model-like gauge group, we find very similar results.
\end{abstract}


\maketitle

\section{Introduction}
Grand unified theories provide an interesting framework for
unification of the strong and electro-weak forces. The minimal simple Lie
group
that could be used to achieve this is $SU(5)$~\cite{orgsu5} or,
as a modification
of this, flipped $SU(5)\times U(1)_X$~\cite{flippedsu5}.
The latter is more interesting from a phenomenological point of view,
because models based on this gauge group might survive the experimental limits
on proton decay.

In this letter we continue with the analysis of~\cite{stat1,stat2}, where
the gauge sector statistics of a type II orientifold has been considered.
Our approach
is inspired by the statistical treatment of the string vacuum problem.
For a recent review on the intersecting brane models we will be dealing with
the reader might want to consult~\cite{isrev}.
We report on a systematic computer analysis of intersecting brane 
models in a $T^6/\bZZ$ orientifold background~\cite{z2xz2} which extends the results
published in~\cite{stat2} in the direction of grand unified theories. We focus on the frequency
distributions of $SU(5)$ as well as flipped $SU(5)$ gauge groups.
Explicit constructions of models of this
type have already been performed in~\cite{su5,moresu5}, non-supersymmetric
models have been
constructed in~\cite{nonsusysu5}.

\section{Setup and Methods}
We work in the intersecting brane picture of type IIA, compactified on a toroidal orientifold of
$T^6/\bZZ$. We use the setup and the notation of~\cite{stat2} and refer the reader to this paper for more details. In particular we are treating only factorisable branes, that can be expressed by their wrapping numbers on the three two-tori of $T^6$. 

The D6-branes wrapping special Lagrangian three-cycles are parametrized by integer-valued
coefficients $X^I,Y^I$, $I\in\{0,\ldots,3\}$. There are two different possibilities for the
geometry of the three $T^2$s, expressed in the three variables
$\beta_i\in\{1,2\}$, $i\in\{1,2,3\}$, where a
value of $2$ stands for a tilted torus. Furthermore we define a rescaling factor
$c:=\prod_{i=1}^3 \beta_i$.

There are three basic constraints to get consistent string vacua in our setup:
\begin{enumerate}
\item The supersymmetry conditions, written in our variables as
\be{eq_susy}
\sum_{I=0}^3\frac{Y^I}{U_I}=0,\quad\sum_{I=0}^3X^IU_I>0.
\ee
They assure that the D-branes wrap special Lagrangian cycles and exclude
the appearance of antibranes.
The $U_I$ parametrise a rescaled version of the complex structure moduli, defined as
$U_I = (U_0, U_i)$ with $U_0 := R^{(1)}_1R^{(2)}_1R^{(3)}_1$ and $U_i := R^{(i)}_1R^{(j)}_2R^{(k)}_2$, where $i,j,k\in\{1,2,3\}$ cyclic and $R^{(i)}_{1/2}$ are the radii of the two-torus $i$.

\item The tadpole cancellation condition for $k$ stacks of $N_a$ branes, given by
\be{eq_tad}
\sum_{a=1}^kN_a\vec{X}_a=\vec{L},
\ee
where the $L^I$ parametrise the orientifold charge. Concretely we have $\vec{L}=\left(8c,\{8\beta_i\}\right)^T$.
\item An additional constraint from K-theory~\cite{ktheory}:
\be{eq_k}
\sum_{a=1}^kN_aY_a^0 \in 2\bZ,\quad
\frac{\beta_i}{c}\sum_{a=1}^kN_aY_a^i \in 2\bZ.
\ee
\end{enumerate}

Chiral matter in a bifundamental representation arises at the intersection
of two stacks of branes with a
multiplicity given by the intersection number
\be{eq_bifundamental}
  I_{ab} = \sum_{I=0}^{3}\left(X_a^IY_b^I-X_b^IY_a^I\right).
\ee
Furthermore, we get symmetric and antisymmetric representations 
\be{eq_symanti}
  \#\Sym_a = \frac{1}{2}(I_{aa'} - I_{a{\rm O}6}),\quad
  \#\Anti_a = \frac{1}{2}(I_{aa'} + I_{a{\rm O}6}).
\ee

In the original $SU(5)$ construction, the standard model particles are embedded in a $\fivebar$ and
a $\ten$ representation of the unified gauge group as follows
\bea{eq_su5embed}\nonumber
SU(5) &\to& SU(3)\times SU(2)\times U(1)_Y,\\\nonumber
\fivebar &\to& (\rep{\bar{3}},\rep{1})_{2/3} + (\rep{1},\rep{2})_{-1},\\
\ten &\to& (\rep{\bar{3}},\rep{1})_{-4/3} + (\rep{3},\rep{2})_{1/3} + (\rep{1},\rep{1})_{2},
\eea
where the hypercharge is generated by the $SU(3)\times SU(2)$-invariant generator
\be{eq_z}
  Z=\mathrm{diag}(-1/3,-1/3,-1/3,1/2,1/2).
\ee

In the flipped $SU(5)$ construction, the embedding is given by
\bea{eq_su5xembed}\nonumber
SU(5) \times U(1)_X &\to& SU(3)\times SU(2) \times U(1)_Y,\\\nonumber
\fivebar_{-3} &\to& (\rep{\bar{3}},\rep{1})_{-4/3} + (\rep{1},\rep{2})_{-1},\\\nonumber
\ten_1 &\to& (\rep{\bar{3}},\rep{1})_{2/3} + (\rep{3},\rep{2})_{1/3} + (\rep{1},\rep{1})_{0},\\
\rep{1}_5 &\to& (1,1)_2,
\eea
including a right-handed neutrino. The hypercharge is in this case given by
the combination
$Y=-\frac{2}{5} Z+\frac{2}{5}X$.

We would like to realise models of both type within our orientifold setup.
The $SU(5)$ case
is simpler, since in principle it requires only two branes, a $U(5)$ brane $a$
and a $U(1)$ brane $b$, which
intersect such that we get the $\fivebar$ representation at the intersection.
The $\ten$ will
be realised as the antisymmetric representation of the $U(5)$ brane.
To get reasonable models,
we have to require that the number of antisymmetric representations is equal
to the number of $\fivebar$ representations,
\be{eq_su5constr}
  I_{ab} = -\#\Anti_a.
\ee

In a pure $SU(5)$ model one should also restrict to configurations with
$\#\Sym_a=0$ to exclude
$\fifteen$ representations from the beginning. It has been proven in~\cite{su5} that in this
case no three generation models can be constructed. Besides,
symmetric representations might also
be interesting from a phenomenological point of view,
thus we will include them in our discussion.

The flipped $SU(5)$ case is a bit more involved since in addition to the constraints of the $SU(5)$
case one has to make sure that the $U(1)_X$ stays massless and the $\fivebar$ and $\ten$ will have the
right charges.
To achieve this, at least one additional brane $c$ is needed. Generically, the
$U(1)_X$ can be constructed as a combination of all $U(1)$s present in the model
\be{u1constrgen}
  U(1)_X=\sum_{i=1}^k x_i U(1)_i.
\ee
The condition that the hypercharge should be massless
can be formulated as
\be{u1massless}
  \sum_{a=1}^k x_aN_a\vec{Y}_a = 0,
\ee
with some unknown coefficients $x_a$.

This condition boils down to a system of linear equations which can be solved
by a standard algorithm. In the case of
models without symmetric representations of $SU(5)$, one can be almost sure
to find a solution, given four or more hidden-sector
brane-stacks. In contrary for models including symmetric representations the
probability for a massless $U(1)$ lies slightly
below 50 percent almost regardless of the number of brane stacks in the hidden
sector.

\section{Results}
Having specified the additional constraints, we use the techniques developed in~\cite{stat2} to
generate as many solutions to the tadpole, supersymmetry and K-theory conditions as possible.
The requirement of a specific set of branes to generate the $SU(5)$ or flipped $SU(5)$ simplifies the
computation and ´gives us the possibility to explore a larger part of the moduli space as compared to
the analysis in~\cite{stat2}.

\begin{figure}[h]
\includegraphics[width=0.9\linewidth]{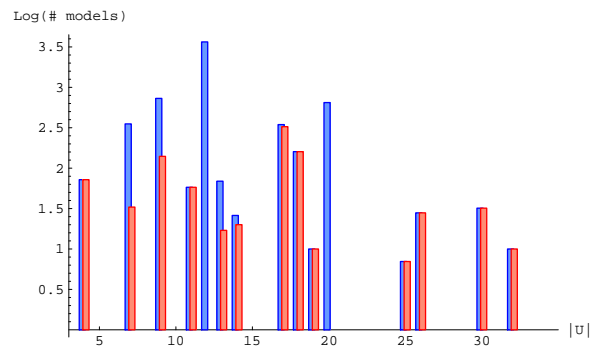}%
\caption{Logarithmic plot of the number of solutions with an $SU(5)$ factor depending on the absolute value of the parameters $U$. The blue bars (left) show the result including models with symmetric representations of $SU(5)$. The red bars (right) represent only solutions without these representations.}%
\label{fig_numsol}
\end{figure}
Before conducting an analysis of the gauge sector properties of the models under
consideration, we would like to check if the number of solutions decreases exponentially for large values\footnote{``Large values'' in our rescaled version of the complex structure parameters means a large difference between the radii of at least one of the three two-tori.}
of the $U_I$. This has been observed in~\cite{stat2} for the general
solutions.
In fig.~\ref{fig_numsol}
the number of solutions with and without
symmetric representations are shown.
The scaling holds in our present case as well, although
the result is a bit obscured by the much smaller statistics.
In total we found 6198 solutions without restrictions on the
number of generations and the presence of symmetric representations.
Excluding these representations reduces the number of solutions to 914.
Looking at the flipped $SU(5)$ models, we found 3816 without the restriction to have a massless $U(1)_X$ and 1970 including this constraint. Imposing the condition to get no symmetric representations reduces the number of solutions further to 394.

\begin{figure}[h]
\includegraphics[width=0.9\linewidth]{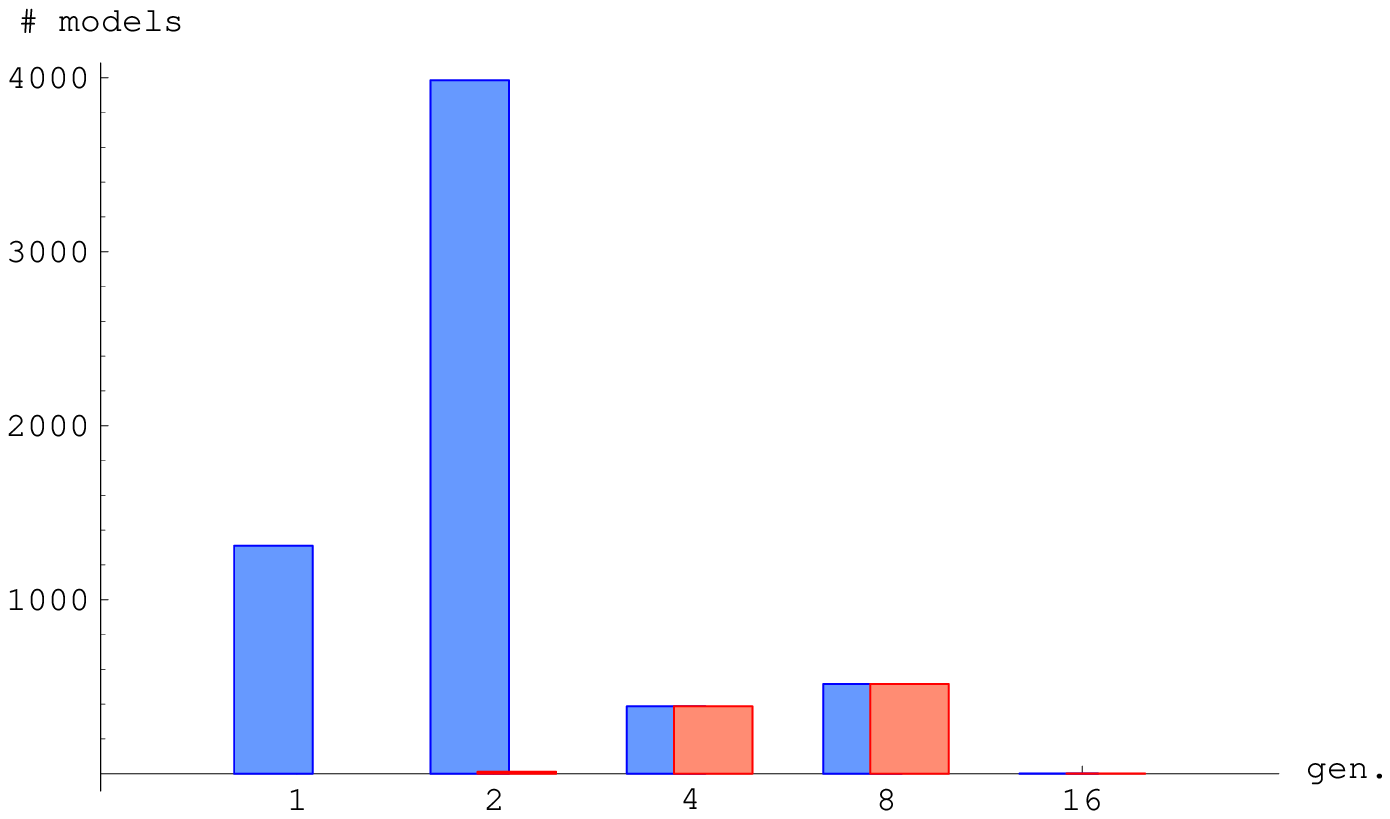}\\
\includegraphics[width=0.9\linewidth]{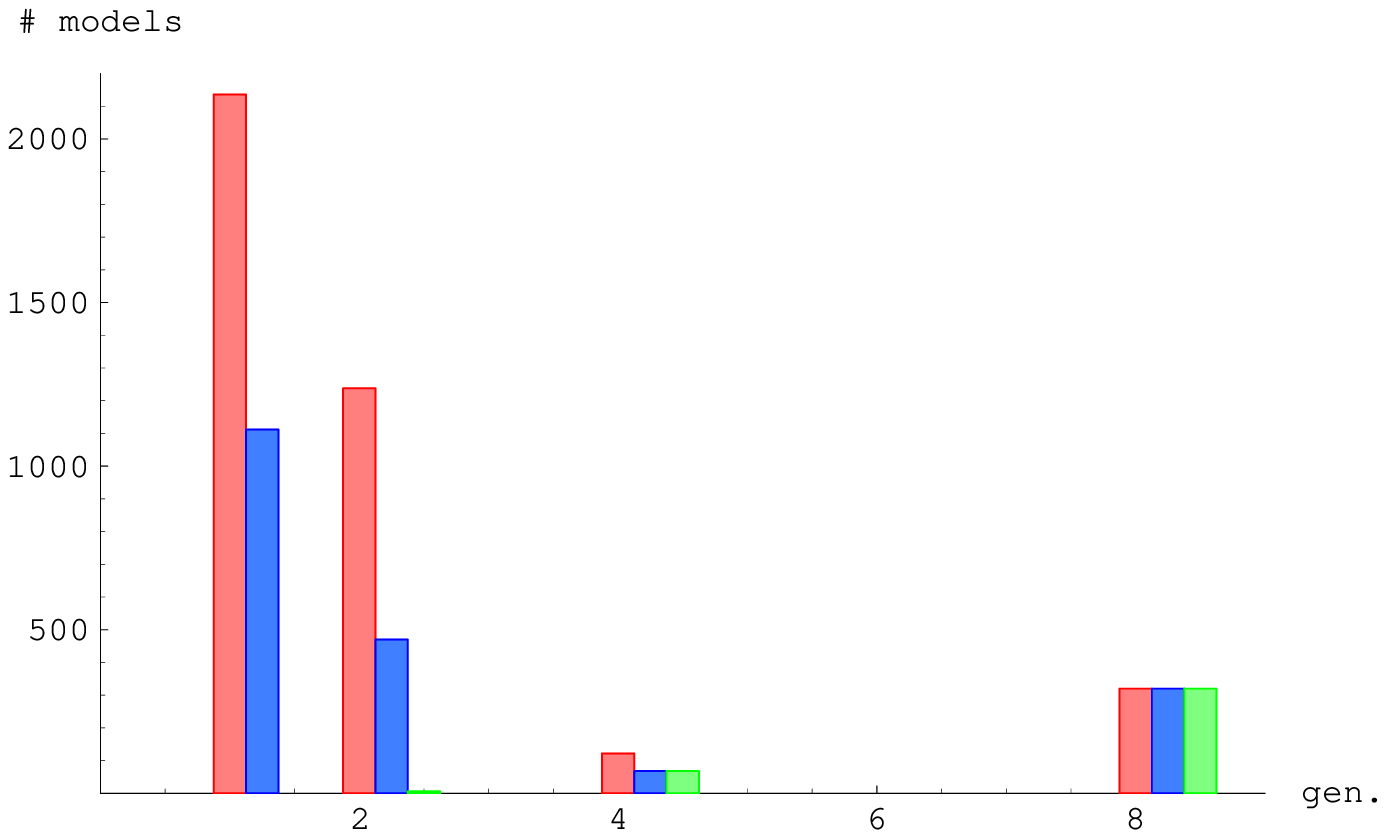}%
\caption{Plots of the number of solutions for different numbers of generations for $SU(5)$ (upper plot): models with (red, left bars) and without (red, right bars) symmetric representations of $SU(5)$; and flipped $SU(5)$ (lower plot): all models (red bars, left), models
permitting a massless U(1) (blue bars, middle) and massless solutions without symmetric representations (green bars, right).}%
\label{fig_gen}
\end{figure}
The correct number of generations turns out to be the strongest constraint
on the statistics in our previous work on standard model constructions.
The $SU(5)$ case is not different in this aspect.
In fig.~\ref{fig_gen}
we show the number of solutions for different numbers
of generations.
We did not find any solutions with three $\fivebar$ and $\ten$
representations. This
situation is very similar to the one we encountered in our previous analysis
of models with a standard model gauge group~\cite{stat2}. An analysis of the
models which have been explicitly constructed showed that they exist only for
very large values of the complex structure parameters. The same is true in
the present case. Because the number of models decreases rapidly for higher
values of the parameters, we can draw the conclusion that these models are
statistically heavily suppressed.

Comparing the standard and the flipped $SU(5)$ construction
the result for models with one generation might be surprising, since there are
more one generation models in the flipped than in the standard case.
This is due to the fact that there are generically different possibilities to
realise the additional $U(1)_X$ factor for one geometrical setup, which we
counted as distinct models.

As in the unflipped case the massless models without symmetric representations have a clear maximum at eight
generations whereas for massless solutions including symmetric representations one or two generations prevail. 
The aforementioned different probability for finding a massless $U(1)$ in the case of models with and without symmetric
representations can also be seen from fig.~\ref{fig_gen}.

\begin{figure}[h]
\includegraphics[width=0.9\linewidth]{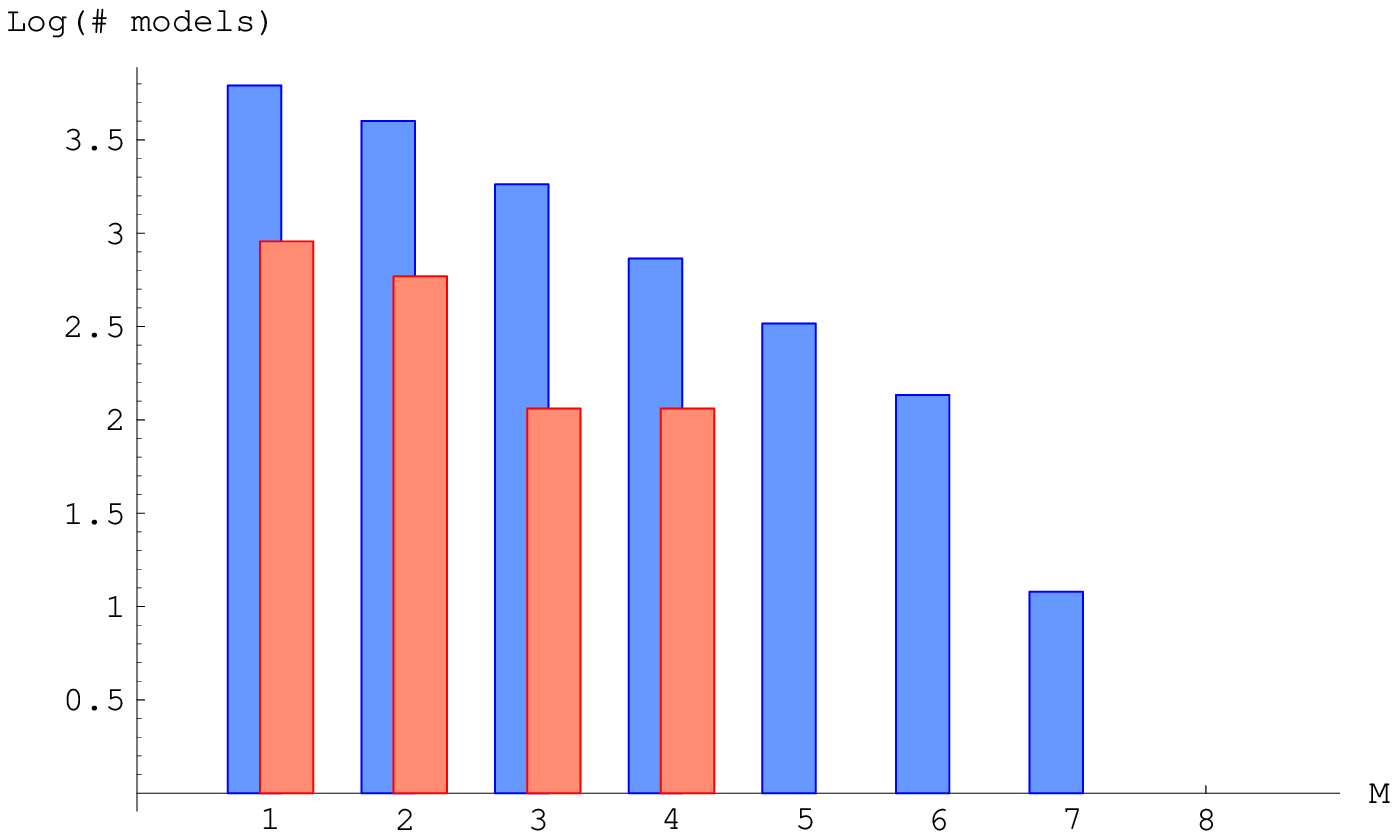}\\
\includegraphics[width=0.9\linewidth]{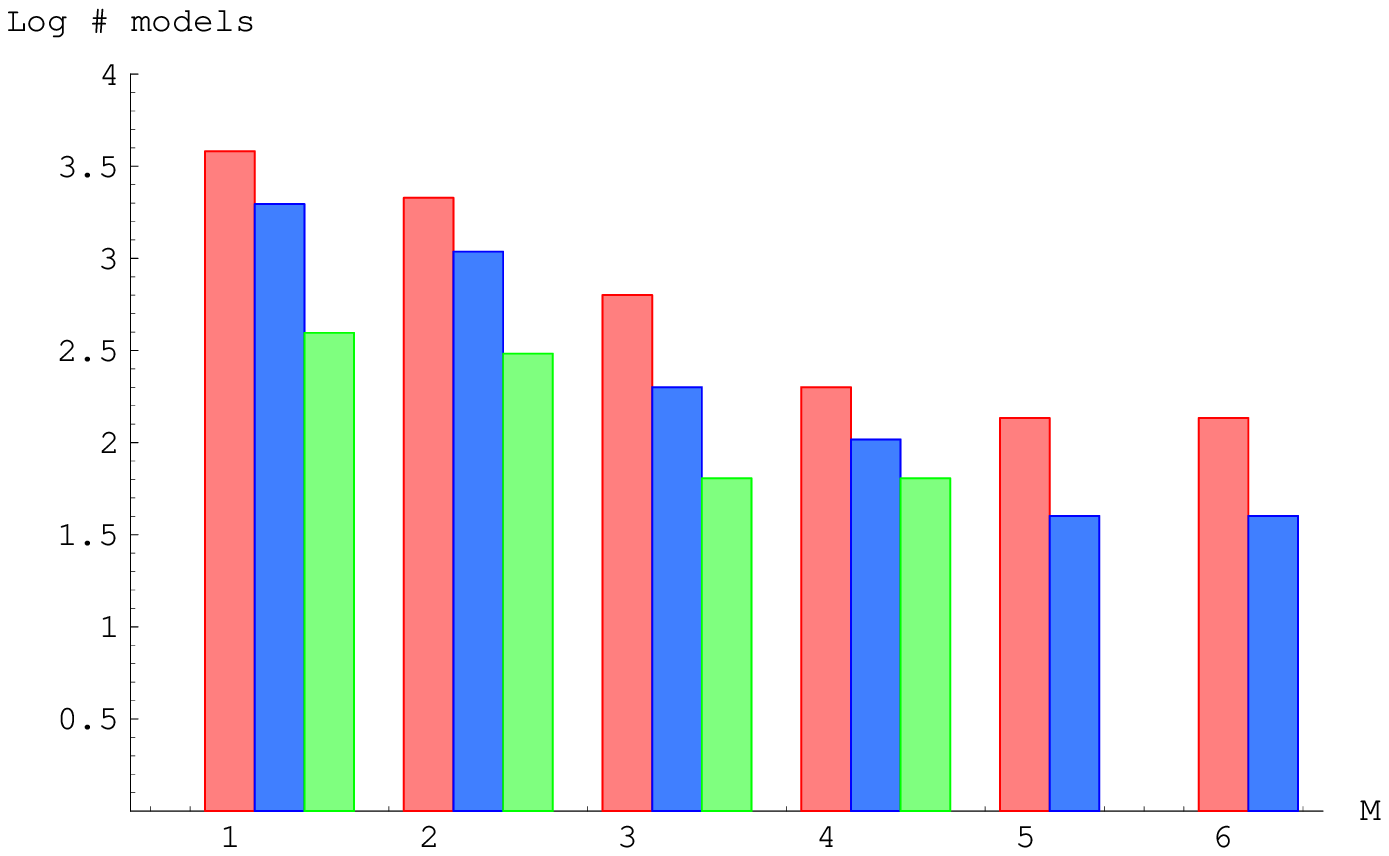}%
\caption{Logarithmic plots of the number of solutions with a specific rank $M$ gauge factor in the hidden sector.
The upper plot shows $SU(5)$, blue (left) and red (right) bars represent solutions with and without symmetric reps. of $SU(5)$; 
the lower plot shows flipped $SU(5)$ models, red bars (left) stand for the total number of solutions, blue (middle) and
green (right) bars represent all massless solutions and those without symmetric reps. of $SU(5)$ respectively.}%
\label{fig_hum}
\end{figure}
Regarding the hidden sector, we found in total only four $SU(5)$ models which
did not have a hidden sector at all - one with 4, two with 8 and one with 16
generations. In the flipped $SU(5)$ case such models do not exist at all.

\begin{figure}[h]
\includegraphics[width=0.9\linewidth]{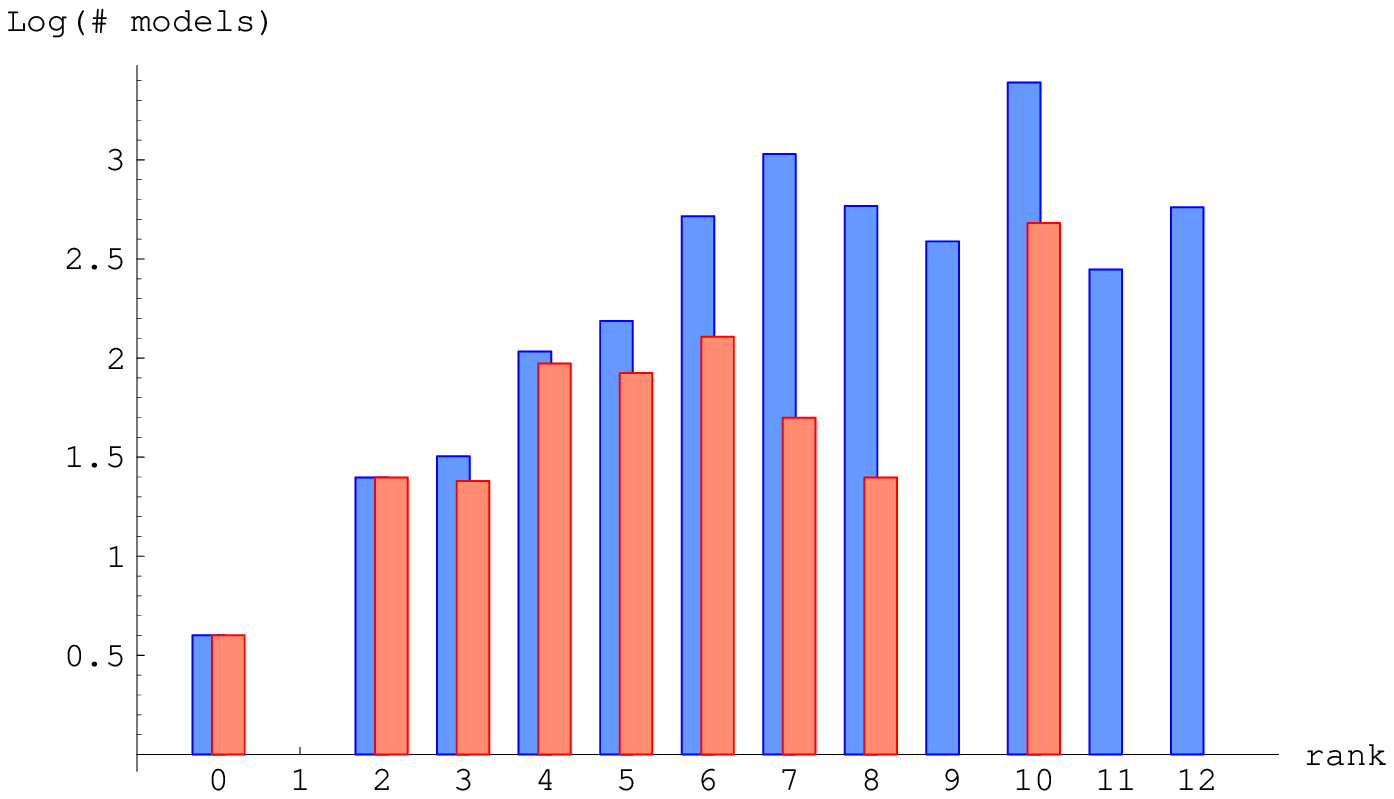}\\
\includegraphics[width=0.9\linewidth]{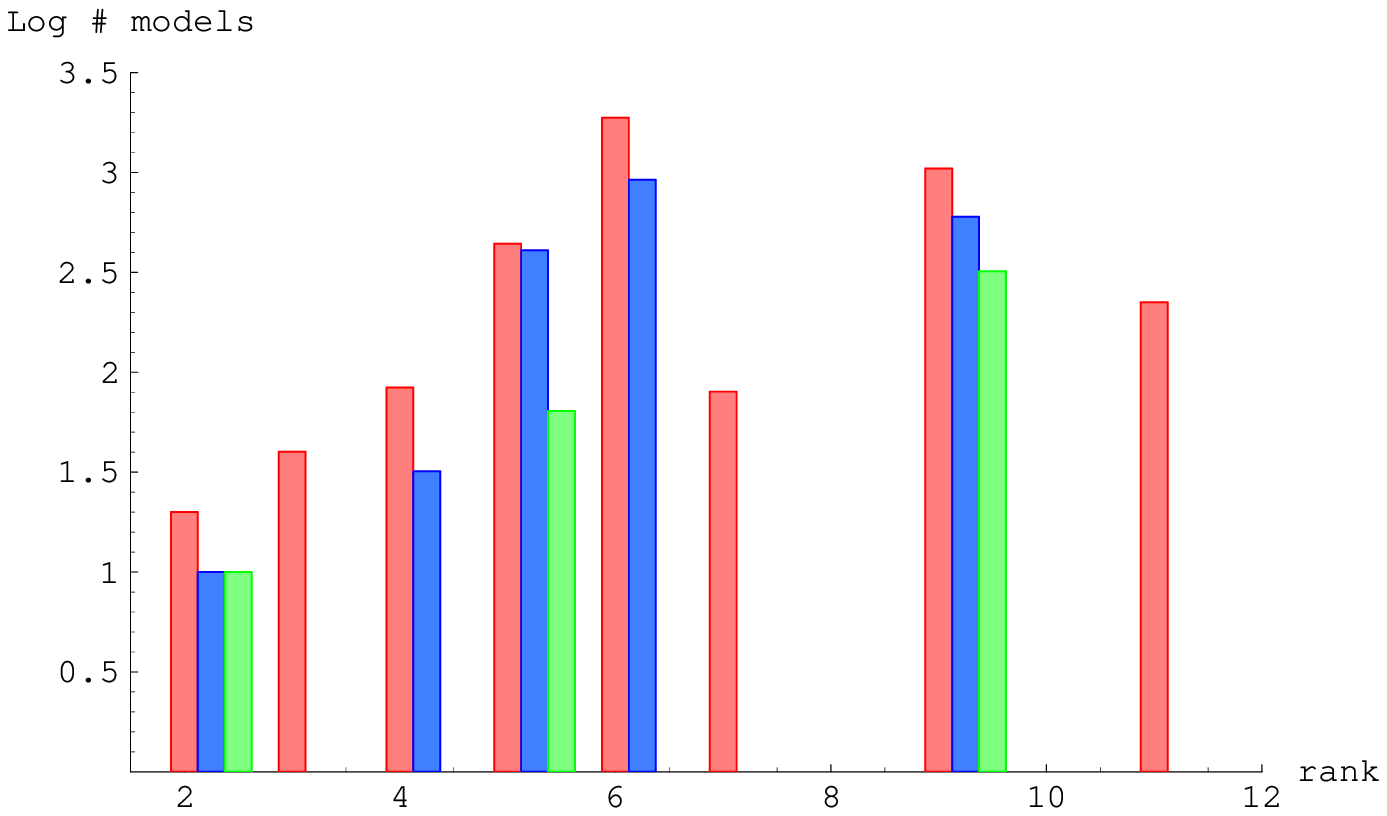}%
\caption{Plots of the number of solutions for given values of the total rank of the hidden sector gauge group.
The upper plot shows $SU(5)$, blue (left) and red (right) bars represent solutions with and without symmetric reps. of $SU(5)$;
the lower one flipped $SU(5)$ models, as before all models (red bars, left), those satisfying the massless condition (blue bars, middle) and 
the massless ones without symmetric representations (green bars, right).}%
\label{fig_hrk}
\end{figure}

The frequency distribution of properties of the hidden sector gauge group,
the probability to find a gauge group of specific rank $M$ and the
distribution of the total rank, are shown in
figs.~\ref{fig_hum} and~\ref{fig_hrk}.
The distribution for individual gauge factors is qualitatively very similar to the one
obtained for all possible solutions in previous work (see figs.~7,~4~resp. of~\cite{stat2}).
This is expected to be the case, since we found in an
earlier analysis that the hidden sector statistics should be generic
and, from a qualitative point of view, independent of the constraints on the
visible sector.
One remarkable difference between standard and flipped $SU(5)$ models
is the lower
probability for higher rank gauge groups.
The massless models show no exceptional behaviour as far as the gauge factors are concerned.

The total rank distribution for both, the standard and the flipped version,
differs in one aspect from the one obtained in~\cite{stat2},
namely in the large fraction of hidden sector groups with a total
rank of 10 or 9, respectively.
This can be explained by just one specific construction which is possible for
various values of the complex structure parameters.
In this setup the hidden sector
branes are all except one on top the orientifold planes on all three tori.
If we exclude this specific feature of the $SU(5)$ construction, the remaining
distribution shows the behaviour estimated from the prior results.

This holds true for the massless models as well, where mainly solutions without symmetric representations
contribute to the peak at a total rank of nine. Yet it is striking that no massless models with a total rank
of three are found and that the massless models without symmetric representations exclusively appear with a rank of two,
five or nine.

Note that while comparing the distributions one has to take
into account that
the total rank of the hidden sector gauge group in the $SU(5)$ case is
lowered by the contribution from the visible sector branes to
the tadpole cancellation conditions. In the flipped case, the additional
$U(1)$-brane contributes as well.

\section{Conclusions}
In this note we presented an analysis of a large number of $SU(5)$ and flipped $SU(5)$ models on a
$T^6/\bZZ$ orientifold.
Our analysis showed that three generation models with a minimal grand unified gauge group
are heavily suppressed in this setup. This result was expected, since we know that the
explicit construction of three generation $SU(5)$ models on this specific orientifold has turned out 
to be difficult. For models without symmetric representations it has
been proven in~\cite{su5} that there exist no models at all.

The analysis of the hidden sector showed that the frequency distributions of the total rank of the
gauge group and of single gauge group factors are quite similar to the results obtained
in~\cite{stat2}. Differences in the qualitative picture result from specific
effects in the $SU(5)$ construction.

Comparing the results for the standard and flipped $SU(5)$ models with and without a massless
$U(1)_X$, we find no
significant differences.
If we allow for symmetric representations, there is basically no additional
suppression factor. If we restrict ourselves to models without these
representations, flipped constructions are three times less likely then the
standard ones.
\vspace{5mm}

\noindent{\bf Acknowledgements}: 

It is a pleasure to thank Carlo Angelantonj, Ralph Blumenhagen, Mirjam Cveti\v{c}, Gabriele Honecker, Tianjun Li, Dieter L\"ust
and Timo Weigand for valuable discussions.

\end{document}